# Digitally reproducing the artistic style of XVI century artist António Campelo in "Alegoria à Prudência"/Allegory of Prudence

João F. Oliveira, João M. Pereira

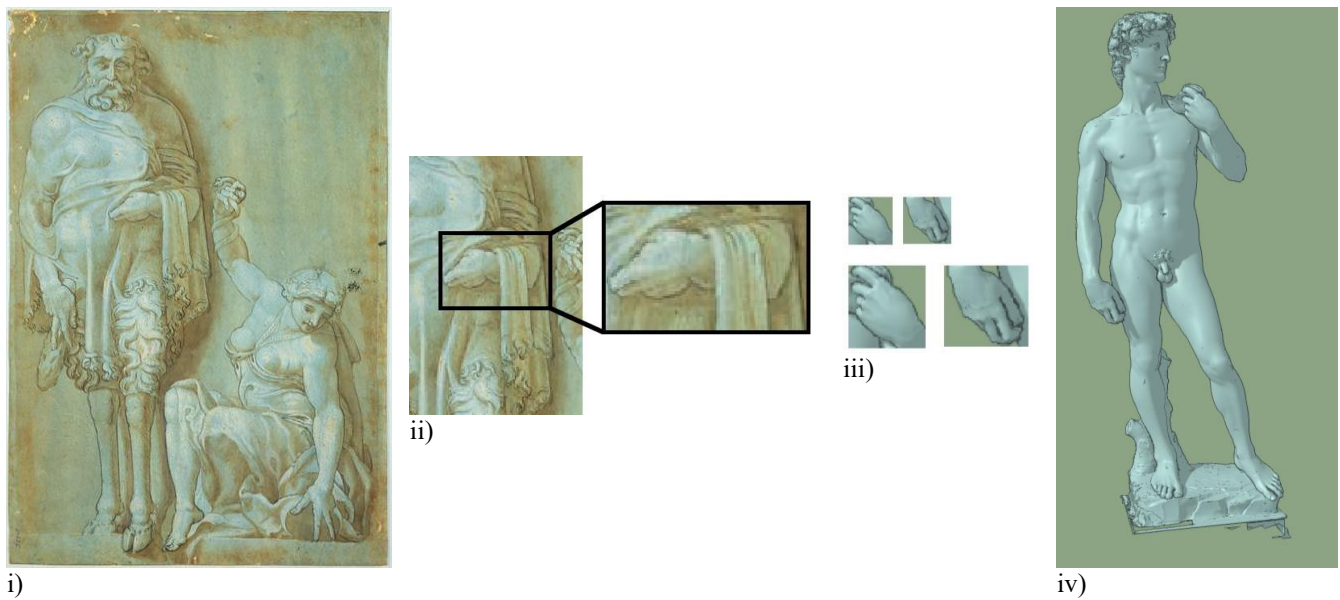

**Figure 1.** i) "Alegoria à Prudência" © IMC/MC. António Campelo´s (XVI century Portuguese artist) 2D illustration; ii)Hand contour line and example insets used for validation in our study(iii)) iv) NPR results of António Campelo style applied to the statue of David.

## Abstract

In this work, the artistic style of the sixteenth century Portuguese artist António Campelo in "Alegoria à Prudência" is analyzed in order to create a computational tool that allows one to transform any 3D digital sculpture model into an image that resembles the modeled style. From this analysis the problem is divided into two parts: detection and drawing of contour lines and the shading of the scene. Several techniques from Non Photorealistic Rendering (NPR) and from Photorealistic Rendering that can resolve the problem are presented and, based on this study, a possible solution is presented. Each modeled rendering component is then analyzed using image based methods against the proposed artistic style and parameters are adjusted for a closer match. In the final stage a group of people was asked to answer a questionnaire where the similarity between the renderings of different objects and the original style was classified according to their personal opinion. One of our findings is that although the source 3D objects cannot be readily found for a direct comparison, nor can the paper medium with centuries old damage be the same, the comparison of sub-parts of both images of the same topology was still possible validating our method and discarding other styles from the comparison.

**CR Categories:** I.3.3 [Computer Graphics]: Picture/Image Generation-Display algorithms; J.5 [Computer Applications]: Arts and Humanities-Fine Arts;

## 1 Introduction

Non-Photorealistic Rendering (NPR) domain includes several subtopics, such as artistic rendering and computational aesthetic, among others. [Battiato et al. 2007] defined this generic domain through its main purpose as follows: to "reproduce aesthetics essence of arts by mean of computational tools". Our work falls into this definition. In particular, it strives to digitally reproduce an artistic style from the sixteenth century Portuguese artist António Campelo in order to understand the intrinsic elements of this style that if modeled adequately could generate an image of a 3D object resembling this artistic style. This artist was chosen because we believe that there is a starching resemblance of some style elements present for example in Fig. 1-i) with some elements that are used in today´s computer graphics, namely shading of 3D laser scanned statues to increase depth perception; and also because he has a predominant place in exhibitions at the Portuguese National Museum of Old Art.

The available António Campelo drawings (Fig. 23, 24, 25) appear to suggest that the artist experimented with different techniques and styles in order to reach the achieved ephemeral contrast between the highlights and shading of Alegoria da Prudência. Namely the highlights in Fig 25 were achieved through a process of oxidation, in Fig. 24 a bi-tonal shading similar to toon shading was used, and finally Fig. 23 uses white gouache on blue paper for this effect. In this work, we focus on capturing the style in Alegoria da Prudência, resorting to the other two works to study and evaluate other style characteristics that are shared for example contour line depiction and shadow projection.

Finally, this selected work of António Campelo (Fig. 1-i)) is thought to have been inspired on a scene painted in 1552 by Italian Mannerist painter Pellegrino Tibaldi on the facade of Vicolo

Savelli in Rome which has since been destroyed but preparatory study sketches apparently still exist [SERRÃO 1991].

## 1.1 Artistic Style Analysis

As mentioned earlier, another drawing from António Campelo that appears to have some of the style characteristics present in Fig. 1-i) is "Sacrifício Pagão"/Pagan sacrifice (Fig. 24, Appendix B), although this drawing does not have the brightness range provided by the white gouache highlights of Fig. 23, we include it here for comparing the other aspects such as contour lines and shadows. From this observation, the problem can be divided, for now, into two parts: the contours creation and the shading and shadowing of the scene.

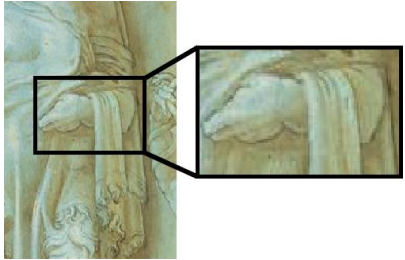

**Figure 2.** Hand contour line detail.

Figure 2 shows how the artist is able to draw a hand with the least amount of lines. These lines delimit the object parts, and give a lot of detail information, like the fingers. But the contours lines are not the only important element here. When there is the need for conveying surface shape and volume, the artist used the shadow and light effect, as can be seen in Fig. 3. The image´s general colour is of the paper/original colour of the canvas, and over that colour the artist applies a white or black ink.

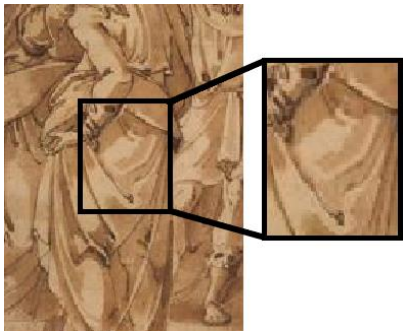

**Figure 3.** Clothing colour detail.

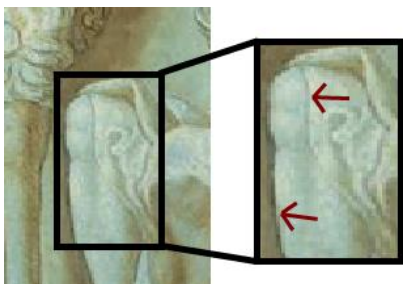

**Figure 4.** Different line style.

### 1.1.1 Contours

Upon a closer examination, it is possible to extract further style elements. The interior contour lines have a different style depending on the depth and shape of the scene. In places where more detail is required, more depth difference with the surrounding area is created, thus the line tone is darker (Fig. 4). The thickness on the other hand, doesn't change significantly, in so much that it can be regarded almost as constant. These lines respect the simplicity of the scene because the scene itself does not have many lines. A balance on detail is present throughout.

### 1.1.2 Colour

Figure 1-i), depicts an almost realistic and naturally lit scene. By changing the paper/canvas colour (Fig. 5) with either a light colour (tint) or dark colour (tone or shade), the physics of light transport and scattering is realistically simulated.

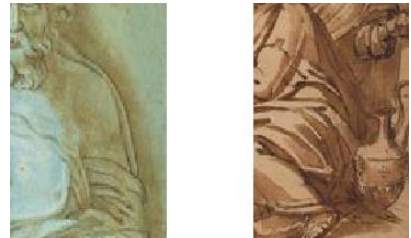

**Figure 5.** coloured regions.

The existing light effect suggests that a directional light source like the sun or moon light was used (Fig. 6), and it is consistent in the whole scene. It is interesting to note, that in the darkest regions, detail was never lost, as the artist was able to darken the regions but at the same time he knew when to stop, thus resulting in dark regions but not too dark, and always less dark than the contour lines.

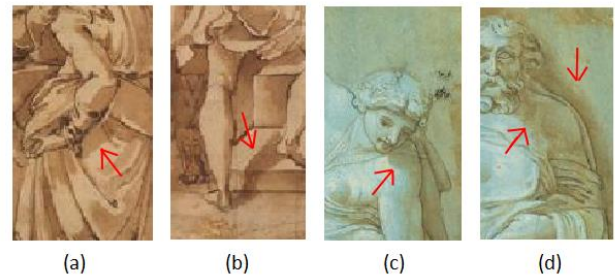

**Figure 6.** Shadow regions.

### 1.1.3 Shadows

Another important visual characteristic present in the work is the existence of shadows that provide the volume information and work together with the light direction. The shadows blend perfectly with the scene. The areas that were darkest do not become darker, and do not have a shadow, as visually that threshold of shape-dark is not crossed. In fact the shadows are so well blended that they are seldom noticed, unless one makes a closer inspection. In addition to shadows that are cast on other objects there are shadows that are cast on the same object. These shadows suggest a soft shadow, with a very small penumbra region.

## 2 Related Work

From the problem resolution standpoint, this section reviews some of the existing techniques that can create an effect similar to the desired ones. These techniques are not restricted to NPR techniques, as the solution requires also photorealistic elements.

One aspect that is particularly challenging in trying to re-create some artistic styles, is that the subjects or objects that were used in the original drawings may no longer exist, or be fictional. Thus style modeling techniques that require the original 3D model [Cole et al. 2008] unfortunately can´t be directly used to learn from the target drawing. Sloan et al. [1999] model the effect of air-brush paintings of human skin, by allowing a user to select areas of scanned art work to create environment maps that are used to produce non-photorealistc images of human characters.

## 2.1 Silhouettes

The silhouette lines of an object are visually the points of the surface that are between the visible and the invisible portion of the surface. This means that in a polygonal model, the silhouette lines are defined as the edges which mark the visible change between a front-facing and a back-facing polygon. This polygon orientation can be easily calculated using the following equation:

$$\vec{n}_{polygon} \cdot (p_{polygon} - p_{camera}) \qquad (1)$$

If the resulting value is greater than zero, the polygon is front-facing, otherwise it is considered to be back-facing.

In a polygonal model there are also other lines to consider: namely ridge and valley lines, border lines and self-intersection lines [Hertzmann and Zorin 2000]. The first lines represent sharp edges that have a greater or smaller angle between two polygons than they would have in a flat surface. Border lines are edges in a model mesh that does not close itself, meaning that the edge only has one adjacent polygon. The last lines exist when two polygons intersect each other.

Existing work in line detection can be divided into two groups, according to the space in which they work: image space algorithms which work on the images resulting from the rendering processing, and the object space algorithms which work directly with the polygonal mesh information. Since our system will use both, we review them in the following section.

### 2.1.1 Image Space Algorithms

The quickest way to detect a silhouette line is through the analysis of the colour buffer. In this buffer, the object´s visible parts will be represented, however this method will also detect false lines based on the shading or the texturing of the object. To address this problem, Saito and Takahashi [1990] suggest the use of the depth buffer and an edge detector such as the Sobel operator. This will improve the quality of the detected lines but unfortunately will still fail depending on the object´s shape. For example, small changes in depth might not be detected and a possible ridge or valley lines will probably be discarded. Hertzmann [1999] and Decaudin [1996] combine the use of the depth buffer with a normal map. In this way, small changes in depth can be detected using the geometry normal vectors information thus generating the best result. These techniques are easy to implement. Their complexity only varies with the size of the input image with which they work, making them a good choice for real time applications. On the other hand, they do not extract lines as individual entities, they only report the places where lines exist, making it impossible to stylize the lines. Another advantage in using these methods is that there are no issues involving the visibility problem, as this has already been taken care of by the software or hardware.

### 2.1.2 Object Space Algorithms

The main problem with image space methods is the impossibility to create a description of the detected lines. Working with the polygonal mesh information will overcome this problem but issues with visibility will arise.

The detection of the silhouette lines is made by calculating which polygonal edges represent the silhouette edges using equation 1, described earlier. This will find all the possible silhouette edges, visible or not. However, because these lines depend on the camera position, each time that position changes the edges need to be calculated again, making this a less attractive method to be used in real time applications.

Buchanan and Sousa [2000] created a technique to accelerate this edge calculation based on a structure named the Edge Buffer. In this structure each edge will have two flags associated: F and B indicating if the edge has an adjacent front-facing or back-facing polygon. For each polygon edge the XOR operator is applied to the respective flags as the polygons have their orientation calculated. Only edges with a FB = 11 will be considered as a silhouette edge. This structure speeds up the edge detection and uses little memory.

Markosian [2000] created a technique based on randomness. Only a portion of the edges is tested. This is based on the premise that only a few edges are silhouette edges and that most of them are connected with each other. If an edge is detected as a silhouette edge, all of the adjacent edges will also be tested while there are more connections to other silhouette edges. Also, the silhouette edges in the previous frame are also tested.

Gooch et al. [1999] and Benichou and Elber [1999] accelerate the edge detection using a Gaussian sphere. In this sphere, an edge is represented as an arc. In the pre-processing phase, all the edges are hierarchically mapped on the sphere with increasing smaller sphere segments. To calculate the edges, the sphere is cut by a plane and the arcs intersected by the cutting plane are the ones representing a silhouette edge. Benichou and Elber also map the Gaussian Sphere onto a cube because the cutting test is more easily carried out using a line rather than an arc.

To detect the border lines, the polygonal mesh can be analyzed and the edges with only one adjacent polygon are marked as a border edge. As for the ridges and valley lines, the angle between both polygons is calculated and tested against a constant that represents the minimal angle that allows an edge to be considered as a ridge or valley edge.

There are other types of line such as the *Suggestive contours* [DeCarlo et al. 2003], *Apparent ridges* [Judd et al. 2007] or *Highlight line*s [DeCarlo et al. 2007]. All these lines create rather nice effects but not the effect that could be used to model the style we wish to recreate.

As mentioned before, these techniques are slower than the image space techniques; however they detect all the silhouette lines and can represent them. Nevertheless after the detection process, it is still necessary to remove the invisible lines or parts of lines.

### 2.1.3 Hidden Line Removal

For the removal of the hidden lines, there are some choices, like the Appel's hidden-line algorithm. However, it would be desirable to use an integrated solution. One example could be to use the graphics API to resolve this problem. The stylized lines could be drawn along with the object and the hidden lines are removed. The problem with this technique is that the lines can be partially removed based on the object geometry. For example, in a valley a line can be partially obscured by the polygons that make the valley walls.

Isenberg et al. [2002] use a depth buffer to test the visibility of the lines iteratively. Breaking the lines into smaller, pixel wide points, every n fragments is pixel depth tested with the average of the surrounding area depth. If the pixel is not visible, the n fragments are discarded.

Markosian [2000] presents a technique that uses an index buffer. In this buffer, every polygon and every edge is drawn with a unique colour representing that particular geometry id. By breaking the lines into smaller fragments and testing the index buffer for the corresponding id in the correct pixel position, he can find out which pieces are visible.

## 2.2 Shading

The object shading is in many cases an important component of a rendered scene. Depending on the intended application, the shading realism level can vary.

Perhaps the most common realistic shading technique is the Phong model. With this model, the colour of a point is calculated based on the ambient, diffuse and specular object colour. In addition the normal of a point is calculated through the linear interpolation of the normals of the polygon vertices. This technique is known to create very good results with little effort.

Gooch et al. [1998] presents Tone Shading. In this technique, the colour of the object is not real, but it can preserve details even in the darkest and brightest areas, thus this is a good technique for technical illustrations where the detail must be preserved. Instead of limiting the colour between white and black, the colour is limited between a warm and a cool colour using a simple colour model equation that takes into account the normal of the point, the camera direction and the chosen cool and warm colour.

Decaudin [1996] created Cartoon Shading. He adapted the Phong model to remove the gradient effect on the objects, making them appear to be shaded with only a few different colours.

## 2.3 Shadows

Shadows are an essential part of a realistic scene. For a computer rendered scene, there are mainly two types of shadow. If a shadow has a constant colour, it is a hard shadow, if however, it has both umbra and penumbra regions, it is said to be a soft shadow. Soft shadows typically are the more realistic of the two.

### 2.3.1 Hard Shadows

Williams [1978] created the Shadow Mapping technique. This technique identifies the part of the scene hidden from a light source, making it possible to create shadows. This technique starts by rendering the scene from the light source position and saving the depth information in a Shadow Map; then when rendering the scene normally, this map is read after the translation of a point position into the shadow map position. If the point depth in the Shadow Map is smaller than the depth to the viewport, there is some geometry between that light source and the point meaning that the point is in the shadow. However, because there are accuracy problems in the depth, the resulting shadow may have artifacts called shadow acne. Another problem with this approach is the size of the Shadow Map. If this size is small, the resulting shadow may have an aliasing effect.

Crow [1977] presents another technique called Shadow Volumes. With this approach, instead of working with images, this technique works with the object´s geometry, consequently it uses more resources than Shadow Mapping. From each light source, every object´s silhouette is expanded to infinity. Inside this volume, every point is in this object´s shadow from a particular light source. Casting a ray from each screen pixel, every time the ray enters a volume, it enters a shadow, and every time the ray exits a volume, it exits a shadow. This process enables one to know whether a point is in the shadow or not. Heidmann [1991] suggests the use of the stencil buffer for this calculation. The shadow volumes are drawn in this buffer two times. In the first time only the front-facing part of the volume is drawn and the value in the stencil buffer is incremented. In the second time only the back-facing part is drawn and the value is decremented. If the resulting value in the stencil buffer is larger than zero, then that pixel is in the shadow.

### 2.3.2 Soft Shadows

One method to create soft shadows is to use the Shadow Mapping from n points slightly translated from the original light source. This means that each map will have the power of (1/n). Joining all the maps together will allow one to make a penumbra region in the edge of the shadow because the sum of powers will be less than 1. Another way to create this shadow is to simply apply any kind of blur effect to the shadow. This requires that the shadow must be drawn first in a single image, blurred, and then applied to the scene.

Reeves et al. [1987] created a technique called Percentage Closer Filtering. In this technique, after creating the Shadow Map, the amount of shadow in a point is based on the adjacent pixels in the map. The number of adjacent points that are in the shadow is counted and the pixel is shadowed accordingly, meaning that if only half of the pixels are in the shadow, the shadow amount for that pixel must be only half of the total shadow.

Fernando [2005] extends this idea and creates the Percentage-Closer Soft Shadows technique that makes softer shadows depend on the object distance. This technique uses a larger or smaller sample zone depending on the distance of the object receiving the shadow to the object blocking the light. The further the blocking object is from the receiver object, the larger the sample zone is and the softer the shadow.

Donnelly and Lauritzen [2006] present the Variance Shadow Maps which calculates how much shadow there is in a point, instead of sampling the surrounding area. The Shadow Map instead of saving only the depth, it saves the depth and the depth multiplied by the depth. By using the Chebyshev's inequality equations and these two values, the quantity of shadow is calculated for each point.

## 3 NPR system of António Campelo's "Alegoria à Prudência"

From the artistic style analysis presented in section 1 and the related work reviewed in section 2, it is possible to address each of the main characteristics of António Campelo´s artistic style used in Fig. 1-i) and to choose the digital technique that best mimics them. The techniques used in our solution were chosen based on the visual-resemblance quality that they provide rather than performance. However the implementation of these techniques enabled us to explore different sampling strategies and achieve a pipeline that is fully automatic, and capable of delivering view-dependent renderings of our modeled rendering style at interactive rates with models of one million triangles.

In this section we present an overview of the main algorithm steps, followed by a section that describes our evaluation strategy, namely how these rendering techniques were evaluated and adjusted during their integration before treating each technique separately and presenting the overall result. Once these parameters are set, the modeled style is ready to be applied to any 3D sculpture model.

## 3.1 Overview

The steps of our algorithm can be viewed in the Fig. 7. Starting with the silhouette lines, those lines in the artistic style show the details of the object very well, up to a point where is possible to distinguish a finger or an eye. Because we need the description of edges in order to change their style, we use a mesh geometry line detection approach in the initial step. However the actual line style is dependent on the surrounding depth and form on the original work, so we use an image-based line detection algorithm as well to create a depth and normal map that we can use to change the style of the other detected lines.

The final lines will be created as the composition of both results, joining both contributions to the final result. For the hidden line removal, the Markosian technique of using an item buffer works well in our application, as the input 3D models that we use in our viewer are typically laser scanned models with a high number of polygons, and as models are supposed to be viewed as a whole, the size of each polygon is quite small, thus there are no noticeable zigzag effects with the edges that would require correction.

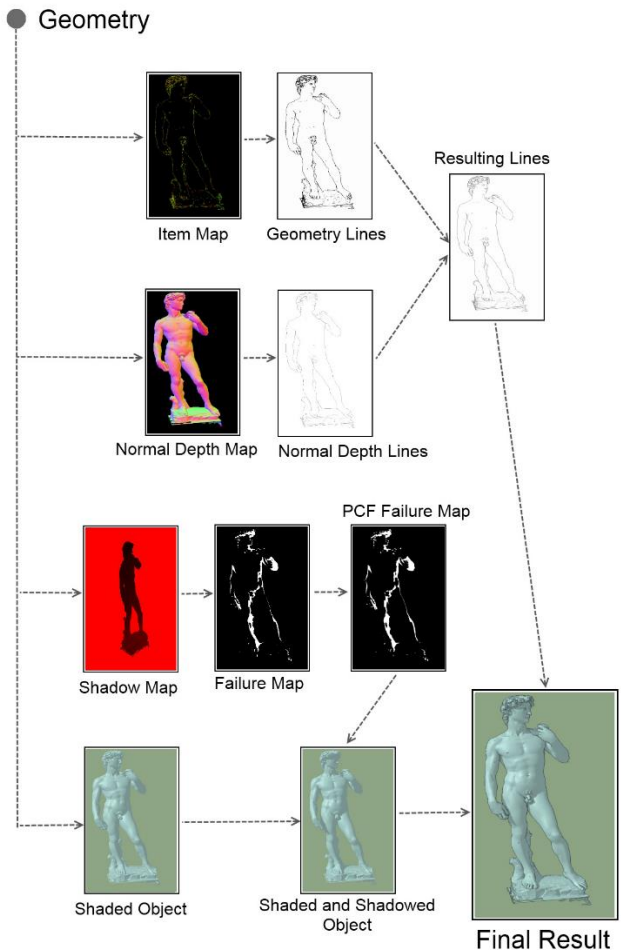

**Figure 7.** Algorithm steps.

The object shading in the original work, as stated before, is realistic, thus the Phong model can be used. However, because there is the need to change the image base colour in intensity, we calculate only the intensity value of the diffuse component, excluding its colour value. The remaining component configurations also need to be adapted, not only in their values, but also in their equation calculation.

As for the shadows, they will need to preserve the ambient colour and need to be relatively soft. The Shadow Mapping technique is a good way to achieve this if combined with the Percentage Closer Filter.

## 3.2 Evaluation

In their paper Isenberg et al. [2006] presented an observational study to examine how people understood hand-drawn pen-and-ink illustrations of objects as compared to NPR depictions of the same 3D shapes.

Hertzmann [2010] and Gatzidis et al. [2008] present various methods for validating an NPR work. Most of these methodologies have two main steps: a mathematical description of the problem if possible, and user tests to validate that description. This evaluation methodology works well for some areas of NPR, for example when an image is changed to give it some extra information and function. In our scenario, the problem of evaluation is the same, if there is no way to formulate a test it is impossible to characterize the style.

Having this in mind, evaluation is performed in intra-generation of the final result. In other words the first step of evaluation is made immediately after a particular technique creates an image. During this process the original drawing image (Fig. 1-i)) is analyzed with image analysis methods and the generated image tested with these values. By dividing the problem into smaller parts, it is possible to test each characteristic and this helps bypass some inherent problems with the analysis. One of the problems is that the scene in the original work, and the 3D models used in the rendering are different, hence results will never be equal. Thus it is necessary to compare similar parts between them.

The non-linear aging of the drawing constitutes also a big problem. With time, the original work loses and changes colour and this has to be taken into account. Finally another problem when comparing a computer generated image and a photo of a drawing is that the medium are different with inherently different characteristics, and if treated as the same the result of the analysis will suffer. There is some NPR work on the imitation of paper and drawing on paper [Sousa and Buchanan 1999; Lee et al. 2007; Wang and Hu 2011], but those techniques do not focus on paper damage due to weathering, which is very pronounced in our study images.

Now that our evaluation strategy has been outlined, we proceed to describe how each component was designed and adjusted to best match the parameters found in the original work (Fig. 1-i)).

## 3.3 Silhouette Lines

The polygonal mesh analysis starts out by finding the front-facing and the back-facing polygons; it then calculates the polygon´s middle position and respective normal along with polygon´s orientation using equation 1 formulated in section 2.1. As for the border lines, these lines can easily be found by checking if each edge has only one adjacent polygon. The ridge and valley lines are detected testing the angle between the edge´s adjacent polygons, using their normals. If this angle is larger than a defined value, the edge is considered a ridge or valley edge.

The next step, is to check which of the detected silhouette lines pass the hidden line removal test. The index buffer is created by rendering all the edges that were detected before; each edge has a unique colour that represents its id. The polygons are then rendered in a black colour and slightly displaced in depth in the direction of the viewport, so that the resulting image will only have the visible lines represented and there are no resulting depth problems from interleaving competing pixels of polygon and edges at the same depth. Each edge is divided into different segments, and its colour identifier is tested against the corresponding colour in the correct index buffer position; only the visible line segments will pass and are considered to be visible silhouette lines.

The Normal Depth Map is created by using a shader. In a single pass, the geometry is rendered and the resulting image colour compacts the normal and depth information in a single RGBA 32 bits coloured pixel. Each pixel has the normal information in the RGB channels and the depth in the A channel. This image is then analyzed using the four corners of each pixel. The differences in the depths and in the normal at each pixel are taken into account to create pixels whose colour only varies between white and black, with black representing the lines with the most differences.

Both of the lines from the geometry analysis and from the normal depth map techniques are then joined together (Fig. 11-e)). Because the normal depth lines are more subtle than others, those lines are important and thus, they need to be almost as visible as they were when detected. As for the geometry analysis lines, those lines are drawn in lighter colour, blurred slightly, and then added to the normal depth lines, making it possible to add the detail to the final lines. The composite image of Fig. 11-e) is composed of 30% geometry analysis lines and 70% normal depth lines respectively. At the same time, this composition strategy allows one to: to add the variation in the lines colour according to geometry; to add the detail transmitted by the geometry analysis lines; and finally, for any error that may exist in either method, to be corrected by the other. In the bottom row of Fig. 11, it is possible to see that the silhouette contour line from the geometry analysis has a gap in David´s under arm due to precision errors (Fig. 11-b)), after the composition with the normal depth lines (Fig. 11-d) the gap is closed (Fig. 11-f)).

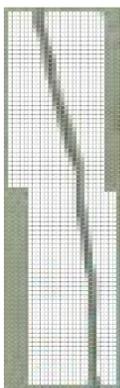
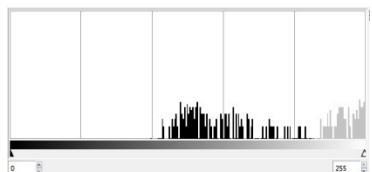

**Figure 9**. The brightness histogram from the previous image (Fig. 8). The grey part to the right is generated by the smooth effect/corner diagonals created during the line extraction process and is ignored.

**Figure 8**. Drawing line thickness with pixels visible.

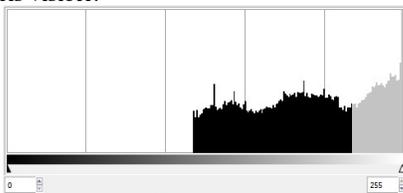

**Figure 10.** The brightness histogram of the generated final lines. Similarly to Figure 9, the grey part to the right, generated during this process, is ignored since it has no impact in the final result.

### 3.3.1 Silhouette Lines Adjustments

The lines present in the original drawing and that were used in the testing were further studied in order to establish their colour and thickness characteristics. By using a standard image editing tool to manually extract the lines (Fig. 8) and calculate the brightness histogram (Fig. 9), it was possible to conclude that the lines, in proportion to the image size, are one or two pixels wide for high contrast lines and smoother lines have a similar width. From the histogram information, it was possible to see that the lines have a colour with brightness values between 0.4 and 0.8 (0 being black, and 1 white). These values were then used in the lines creation process (Fig. 10).

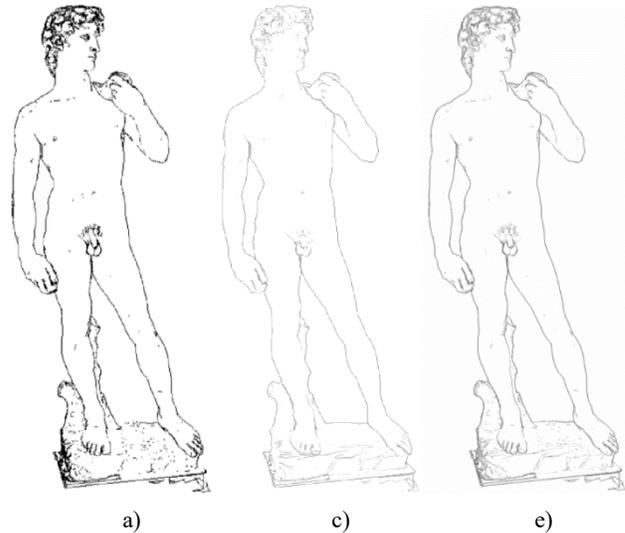

a)   c)   e)

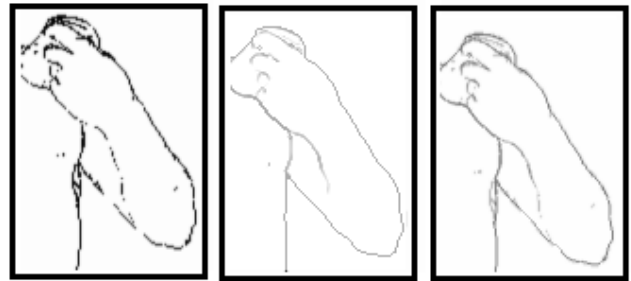

b)   d)   f)

**Figure 11**. Rendered line composition – a)&b): Lines from the geometry analysis; c)&d): Lines from the normal depth map; e)&f) Final lines (composite image).

### 3.4 Shading

The Phong model is used almost identically as defined. The diffuse component colour is removed in order to simulate the changes of the illumination in a paper, with the use of a white and a black colour. Hence, the final pixel colour is the sum between the ambient colour, the diffuse intensity and the specular contribution.

### 3.4.1 Shading Adjustments

For the adjustments made in the illumination, the minimal colour brightness was calculated using the darkest places in the original drawing (with a brightness of 0.55 of white). This colour is then used as the ambient component colour.

To model the light source position, some guide lines (Fig. 12) were created connecting some of the shadows extreme boundary points to the part of the object that created that particular shadow point (Fig. 13). The angle between these guide lines and the horizontal image plane was calculated and averaged resulting in a 45 degree angle. But this only represents a 2D space. For the third dimension, the shadows of the drawing were compared to the part of the object that created them. Both shadow size and object size are identical, meaning that not only the light source makes 45 degrees with the horizontal plane, the light source creates shadows identical in size to the object parts.

As for the light intensity, first various histograms from the lighter places were created. Those histograms revealed that most of the brightness was between 0.6 and 0.8 of white. These intensity ranges represent the values allowed in the lightest brightness regions. Secondly, the size of the specular areas was calculated and used to adjust both the diffuse intensity and the specular contribution.

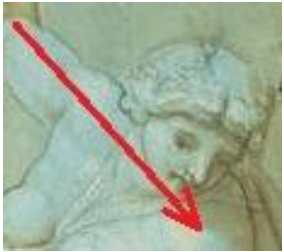 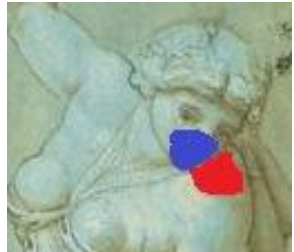

**Figure 12.** One of the light direction guide lines.  **Figure 13.** The part of the object in blue and the casted shadow in red.

### 3.5 Shadows

The Shadow Map is created using the light source chosen before. It uses an orthographic projection to simulate a directional light like the Sun or Moon light. The scene depth is stored in a texture with each pixel corresponding to a single red channel with 32 bits and the texture size is double the size of the viewport in both directions for more precision in the casting of the shadows.

The scene is now rendered once again and the depth test is made between the scene depth and the value stored in the Shadow Map; the result of each pixel test is saved in a Failure Map, which represents the pixels where a shadow exists.

For the depth test, in order to address precisions problems, a bias value is used instead, rather than making the test directly with the raw depth data. Finally, the Percentage Closer Filter technique is applied to the Failure Map, this map represents how much shadow there is in a single pixel, instead of reporting the binary value of "this pixel" is "in" or "out" of the shadow.

The last step is to apply shadows to the shaded image. For this, the modified Failure Map is read. The value is saved in this map and has a value between zero and one, with the zero value corresponding to a pixel fully illuminated and the value of one to correspond to a point in the shadow. The final pixel colour is calculated by the interpolation between the pixel shaded colour (fully illuminated colour) and the ambient colour (shadow colour) based on the value in the map. Figure 14 shows a preliminary result.

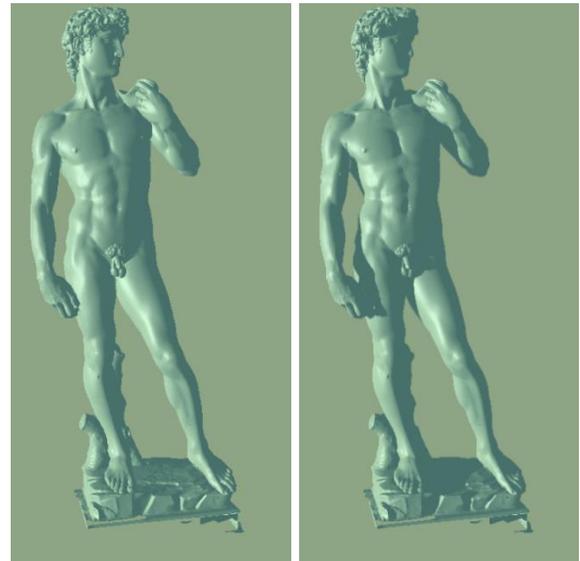

**Figure 14.** David - left: Shaded, right: Shaded and shadowed.

### 3.5.1 Shadows Adjustments

At this point there is only one adjustment left to make. The minimal ambient colour is maintained by the process but the light source position has to be changed to ensure that the third dimension creates a shadow with a similar size to the object part that cast it. Hence we find this position manually. Figure 15 shows the adjusted shaded and shadowed results.

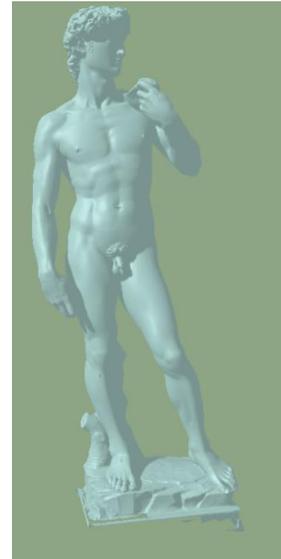

**Figure 15.** Shaded and shadowed David.

### 3.6 Results

The final image results from the darkening of the shaded and shadowed image by the value of the generated silhouette lines. Because these lines have a value between 0 and 1, the final result will maintain the colour characteristics of this artistic style. The image based analysis made to both the result and the original drawing image tried to divide the problem into smaller testable characteristics. With this approach it is easier to compare and prove that both images are similar, and at the same time it was also possible to ignore problems with the paper quality and age of the drawing. By proving the resemblance of each part of the image individually (Fig. 16) it becomes easier to understand how good the final result is as a whole. That final image result is the goal of this work, so there is no attempt to mimic the smaller parts in particular, rather we are interested to mimic just the characteristics that all those parts share. The result of our algorithm can be seen with the rendering of the David statue in the (Fig. 1- iv)). The other results shown in the appendix A were generated using the same configuration parameters used for the David statue (Fig. 1 -iv)), hence no extra adjustments were made.

## 4 Questionnaire

Because art is not discrete, the opinion of people is the reason the art is art. To validate the images generated before, a group of 20 people were asked to answer a questionnaire. In this section we detail the methodology of our evaluation, the questionnaire and its objective and analyze the collected results.

### 4.1 Evaluation methodology

Seven men and thirteen women were asked to answer a questionnaire. The participants were aged from twenty four to seventy four years old and six of them were considered to be art lovers (these people did not necessarily hold degrees in art, but had experience in art, created paintings themselves for example). Five of the participants were students, four were employed, and eleven, including the art lovers, were retired. The test took place in a dedicated room without exterior or interior interferences. Each session of tests was twenty minutes long, with a five minutes briefing dedicated to introduce the work and describe the objective of the test and the nature of questions. The remaining fifteen minutes were for the test itself, the participant answering the questionnaire. The room was well lit, and all the materials used were based on paper, without any electronic device. The answers were anonymous, and there is no knowledge of which answers corresponded to the people linked to art.
In the introduction the objective of the work and the structure of the test was explained. The most important aspect of this explanation was to make it clear that the images that were going to be presented did not represent the same scene nor did they try to do so. Instead it was the underlying artistic style that was being assessed. It was also important to explain that it was necessary to bear in mind that there was no attempt to model the drawing, the age or damage on the paper or canvas.
The questionnaire consisted of a sequence of twenty eight questions, in which two printed images were presented side by side in turn. The first image (the image to the left) corresponded to an image or part of an image that was generated by our system and the second image (the image to the right) corresponded to the respective part from the available drawings with the artistic style being analyzed. In each question, it was asked for a person to give a value from zero to ten for a given characteristic in the images, in which ten indicated that the characteristic was completely achieved, while a classification of zero meant that it failed completely. It was also asked to write a brief explanation or comment their voting whenever they thought it was necessary.

### 4.2 Questionnaire and objective

In general the aim of the questionnaire was to evaluate all the analysis, modeling and parameter adjustments that were carried out to obtain the final image result for the statue of David (Fig.1-iv). After a first sequence of ten questions comparing the specific style characteristics in the various parts of this image against corresponding parts of Fig. 23, a sequence of eighteen comparisons between whole image results using the same NPR settings with other 3D models and the other available drawings from the artist (Appendix A) were presented for cross-referencing. By structuring the questionnaire into these two sequences, the parameter adjustment process and the evaluation work can be directly complemented and completed. This methodology allows that the results in a general setting can be properly analysed. Admittedly the second sequence could have been more exhaustive, analyzing each image by sub-parts and as a whole, but this would make the questionnaire too long and thus tiring. Instead the focus was to complete the study and at the same time analyze other results with the same rules. For questions one to five, subjects were asked to evaluate only the shape, colour and detail of contour lines.

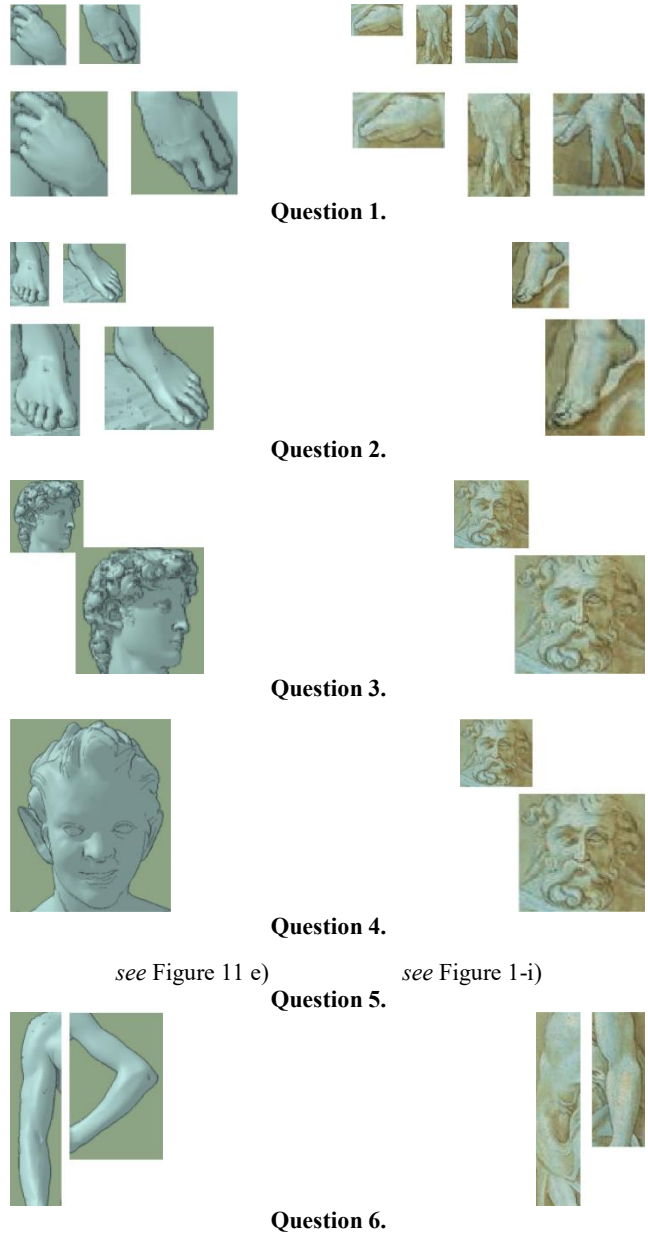

**Question 1.**

**Question 2.**

**Question 3.**

**Question 4.**

*see* Figure 11 e)     *see* Figure 1-i)
**Question 5.**

**Question 6.**

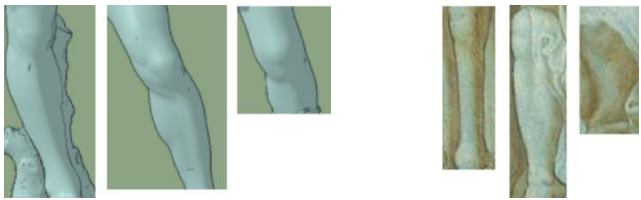
**Question 7.**

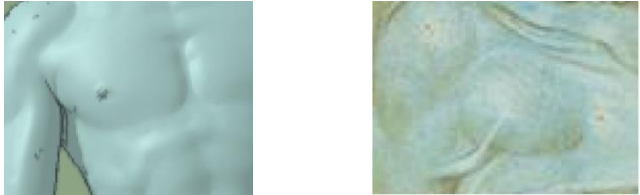
**Question 8.**

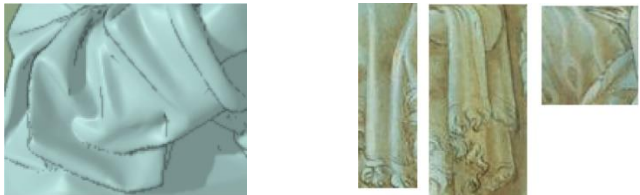
**Question 9.**

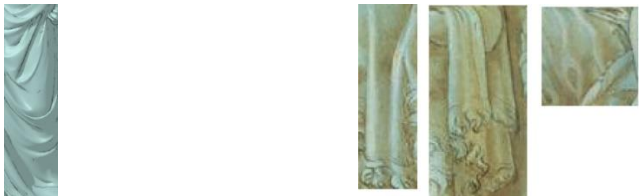
**Question 10.**

**Figure 16.** Image pairs used in the questionnaire referring to the first 10 questions (left: NPR, right:left).

For question one, two and three, where the results on the hands, feet and head respectively were asked to be compared, it is expected to find results classified with a high similarity since these sub-parts were directly evaluated with image-based methods during the parameter adjustment phase. Regarding question four, a different head (Fig. 18), with less detail but generated with the same parameters, was used in the comparison, it is expected to obtain a medium match on similarity as the modeled detail level in the 3D model is different than the detail level in the drawing. In question five, an assessment of the same contour line characteristics is asked to be made, but in a general setting, comparing the whole of Fig.11-e) and Fig. 23, it is expected to find a high similarity match since the the sub-parts of the image were directly used in the evaluation and parameter adjustment phase.

Concerning questions six to ten, subjects were asked to evaluate the similarity of illumination and shadows. In question six, subjects were asked to compare the results on the arms of the David (Fig. 1-iv) and Fig. 19) with the left and right arms present in Fig. 23. Since the images represent different scenes, and the differences due to the age of the paper are more noticeable in smaller images, it is expected that the illumination will not be entirely similar, as different peaks in illumination exist. A medium to high classification is expected in question six (arms), seven (legs) and eight (chest) for the same reasons.

Although the shading and shadows of draping was not specifically modeled, we included question nine (Fig. 20) and ten (Fig. 21) to see if the style effect was achieved in the presence of more complex elements. As with questions six to eight, a medium to high classification is expected for the same reasons. Table 1 summarizes the expected results of the first sequence of questions.

**Table 1.** Expected classification from questions one to ten; key: H(High), MH(Medium-High), M (Medium), ML(Medium-Low)

|  | 1 | 2 | 3 | 4 | 5 | 6 | 7 | 8 | 9 | 10 |
|---|---|---|---|---|---|---|---|---|---|---|
| Lines Test | x | x | x | x | x |  |  |  |  |  |
| Illumination Test |  |  |  |  |  | x | x | x | x | x |
| Expected classification | H | H | H | M | H | MH | MH | MH | MH | MH |

For questions eleven to twenty eight, subjects were asked to evaluate alternately the characteristics of the lines and illumination between three generated whole images using the same parameters and three drawings from the artist (Table 2). The expected classification for each of these questions follows a constant order. For questions related to Figure 24 (13, 14, 19, 20, 25 and 26), since the differences in illumination are large (note the completely different shadow orientation on the left leg of the center figure of Fig. 24, and the reduced gradient in the shaded results when compared to Fig. 23) it is expected that the illumination test will receive a bad voting and that this will also affect somewhat the classification of the line contour test. Question eleven and twelve address results using the David statue model (Fig. 1-iv), hence a high classification is expected. In the case of question fifteen and sixteen since the shadow orientation and gradient range of the shading is more similar to the modeled style in Fig. 23, a medium-high classification is expected. In the remaining questions it is expected to obtain a classification always slightly lower than the analogous question (seventeen and twenty three when compared to eleven, eighteen and twenty four when compared to twelve, twenty one and twenty seven when compared to fifteen, finally twenty two and twenty eight when compared to sixteen). Question eleven is a control question, since it is the same question as question five. Both address the contour line characteristics using whole images, however question eleven includes all the other style characteristics. If a large difference exists between the classification of question five and eleven, then this suggests that the subject does not keep a consistent criteria or is not being very attentive in the answers, question eleven also enables one to see how the final composition between contour lines, illumination and shadows is regarded.

**Table 2.** Expected classification and test image sequence for questions eleven to twenty eight; key: H(High), MH(Medium-High), M (Medium), ML(Medium-Low)

|  | 11 | 12 | 13 | 14 | 15 | 16 | 17 | 18 | 19 | 20 | 21 | 22 | 23 | 24 | 25 | 26 | 27 | 28 |
|---|---|---|---|---|---|---|---|---|---|---|---|---|---|---|---|---|---|---|
| Fig 1-iv | x | x | x | x | x | x |  |  |  |  |  |  |  |  |  |  |  |  |
| Fig 18 |  |  |  |  |  |  | x | x | x | x | x | x |  |  |  |  |  |  |
| Fig 19 |  |  |  |  |  |  |  |  |  |  |  |  | x | x | x | x | x | x |
| Fig 23 | x | x |  |  |  |  | x | x |  |  |  |  | x | x |  |  |  |  |
| Fig 24 |  |  | x | x |  |  |  |  | x | x |  |  |  |  | x | x |  |  |
| Fig 25 |  |  |  |  | x | x |  |  |  |  | x | x |  |  |  |  | x | x |
| Lines Test | x |  | x |  | x |  | x |  | x |  | x |  | x |  | x |  | x |  |
| Illumination Test |  | x |  | x |  | x |  | x |  | x |  | x |  | x |  | x |  | x |
| Expected classification | H | H | M | ML | MH | MH | MH | MH | M | ML | M | M | MH | MH | M | ML | M | M |

### 4.3 Results

The next table presents the voting made by each participant/subject of the questionnaire (Table 3). The average voting of each question (Table 4) as well as the average vote per subject was computed to help analysis (Table 5).

**Table 3.** Answer to the questionnaires.

| Subject | 1 | 2 | 3 | 4 | 5 | 6 | 7 | 8 | 9 | 10 | 11 | 12 | 13 | 14 | 15 | 16 | 17 | 18 | 19 | 20 | 21 | 22 | 23 | 24 | 25 | 26 | 27 | 28 |
|---|---|---|---|---|---|---|---|---|---|---|---|---|---|---|---|---|---|---|---|---|---|---|---|---|---|---|---|---|
| Subject #1 | 7 | 5 | 5 | 8 | 8 | 4 | 6 | 8 | 7 | 6 | 6 | 7 | 5 | 8 | 6 | 8 | 5 | 3 | 7 | 6 | 3 | 3 | 4 | 5 | 5 | 7 | 7 | |
| Subject #2 | 8 | 8 | 8 | 4 | 5 | 6 | 7 | 7 | 7 | 9 | 8 | 9 | 2 | 6 | 4 | 4 | 7 | 4 | 5 | 4 | 6 | 8 | 7 | 4 | 4 | 7 | 6 | |
| Subject #3 | 8 | 7 | 5 | 6 | 5 | 5 | 6 | 5 | 7 | 5 | 5 | 5 | 6 | 3 | 5 | 4 | 6 | 9 | 3 | 8 | 3 | 4 | 5 | 8 | 3 | 7 | 5 | 7 |
| Subject #4 | 7 | 7 | 7 | 9 | 10 | 4 | 7 | 6 | 5 | 5 | 8 | 8 | 9 | 3 | 10 | 9 | 8 | 7 | 4 | 9 | 7 | 7 | 7 | 8 | 7 | 9 | 8 | 8 |
| Subject #5 | 5 | 7 | 4 | 3 | 5 | 4 | 7 | 7 | 4 | 7 | 3 | 7 | 4 | 3 | 7 | 7 | 8 | 5 | 7 | 5 | 6 | 8 | 2 | 7 | 4 | 2 | 4 | 7 |
| Subject #6 | 6 | 5 | 6 | 2 | 9 | 3 | 7 | 7 | 5 | 5 | 5 | 5 | 5 | 4 | 5 | 5 | 5 | 6 | 2 | 5 | 4 | 4 | 7 | 4 | 5 | 6 | 2 | 5 |
| Subject #7 | 8 | 7 | 5 | 4 | 8 | 5 | 6 | 6 | 5 | 6 | 7 | 7 | 8 | 5 | 7 | 5 | 6 | 8 | 4 | 5 | 5 | 5 | 8 | 6 | 5 | 5 | 6 | 7 |
| Subject #8 | 8 | 8 | 7 | 5 | 7 | 6 | 7 | 7 | 5 | 6 | 7 | 8 | 6 | 5 | 7 | 6 | 8 | 9 | 5 | 8 | 6 | 7 | 8 | 7 | 5 | 6 | 7 | 8 |
| Subject #9 | 7 | 7 | 8 | 6 | 6 | 6 | 7 | 7 | 6 | 5 | 6 | 6 | 5 | 7 | 7 | 7 | 6 | 7 | 5 | 5 | 5 | 6 | 7 | 7 | 5 | 5 | 6 | |
| Subject #10 | 6 | 7 | 7 | 5 | 7 | 5 | 6 | 5 | 7 | 7 | 8 | 7 | 8 | 5 | 8 | 8 | 8 | 8 | 5 | 7 | 7 | 7 | 9 | 7 | 6 | 7 | 6 | 6 |
| Subject #11 | 4 | 5 | 5 | 6 | 5 | 5 | 4 | 3 | 5 | 5 | 5 | 6 | 7 | 3 | 7 | 5 | 5 | 7 | 5 | 6 | 5 | 5 | 6 | 6 | 4 | 6 | 5 | 7 |
| Subject #12 | 8 | 7 | 7 | 5 | 8 | 6 | 7 | 8 | 6 | 8 | 5 | 5 | 9 | 9 | 9 | 6 | 7 | 4 | 5 | 6 | 4 | 4 | 5 | 4 | 4 | 5 | 5 | 5 |
| Subject #13 | 5 | 6 | 7 | 4 | 6 | 4 | 7 | 5 | 4 | 5 | 4 | 5 | 6 | 5 | 6 | 3 | 5 | 4 | 5 | 5 | 7 | 6 | 7 | 6 | 7 | | | |
| Subject #14 | 6 | 5 | 6 | 5 | 7 | 5 | 6 | 6 | 5 | 5 | 6 | 7 | 6 | 4 | 7 | 7 | 7 | 8 | 6 | 7 | 6 | 6 | 6 | 8 | 5 | 6 | 6 | 6 |
| Subject #15 | 7 | 6 | 5 | 8 | 7 | 7 | 5 | 8 | 7 | 7 | 5 | 7 | 7 | 5 | 8 | 7 | 7 | 8 | 5 | 7 | 7 | 8 | 7 | 7 | 7 | 6 | 8 | |
| Subject #16 | 9 | 7 | 8 | 4 | 9 | 6 | 7 | 8 | 5 | 7 | 9 | 8 | 8 | 5 | 6 | 6 | 9 | 7 | 6 | 4 | 5 | 4 | 4 | 3 | 4 | 4 | 5 | |
| Subject #17 | 8 | 8 | 7 | 4 | 6 | 4 | 5 | 4 | 6 | 4 | 8 | 6 | 8 | 3 | 8 | 5 | 4 | 5 | 3 | 5 | 5 | 5 | 6 | 6 | 4 | 4 | 3 | 5 |
| Subject #18 | 7 | 7 | 8 | 6 | 7 | 5 | 7 | 7 | 7 | 7 | 8 | 8 | 5 | 9 | 8 | 7 | 8 | 5 | 7 | 6 | 5 | 5 | 5 | 6 | 6 | 5 | 5 | |
| Subject #19 | 6 | 6 | 5 | 5 | 6 | 4 | 5 | 6 | 4 | 5 | 6 | 7 | 7 | 4 | 6 | 4 | 5 | 5 | 2 | 4 | 3 | 4 | 5 | 5 | 4 | 5 | 4 | 6 |
| Subject #20 | 7 | 7 | 6 | 5 | 8 | 5 | 7 | 7 | 5 | 5 | 5 | 6 | 6 | 5 | 7 | 7 | 6 | 7 | 7 | 7 | 6 | 8 | 7 | 6 | 6 | 7 | | |

**Table 4.** Average voting per question.

| | 1 | 2 | 3 | 4 | 5 | 6 | 7 | 8 | 9 | 10 | 11 | 12 | 13 | 14 | 15 | 16 | 17 | 18 | 19 | 20 | 21 | 22 | 23 | 24 | 25 | 26 | 27 | 28 |
|---|---|---|---|---|---|---|---|---|---|---|---|---|---|---|---|---|---|---|---|---|---|---|---|---|---|---|---|---|
| Average | 6,85 | 6,6 | 6,3 | 5,05 | 6,95 | 4,95 | 6,3 | 6,35 | 5,6 | 5,7 | 6,2 | 6,8 | 7,05 | 4,2 | 7 | 5,95 | 6,2 | 7,3 | 4,4 | 6,1 | 5,25 | 5,5 | 6 | 6,3 | 4,9 | 5,7 | 5,25 | 6,5 |

**Table 5.** Average voting per subject.

| | S#1 | S#2 | S#3 | S#4 | S#5 | S#6 | S#7 | S#8 | S#9 | S#10 | S#11 | S#12 | S#13 | S#14 | S#15 | S#16 | S#17 | S#18 | S#19 | S#20 |
|---|---|---|---|---|---|---|---|---|---|---|---|---|---|---|---|---|---|---|---|---|
| Average | 5,96 | 6,11 | 5,54 | 7,25 | 5,36 | 4,96 | 6,04 | 6,75 | 6,14 | 6,75 | 5,14 | 5,93 | 5,32 | 6,04 | 6,82 | 6,14 | 5,32 | 6,57 | 4,93 | 6,39 |

The first step in analyzing the information is to define what is regarded as a high, medium and low voting. A starting point would be to use the intervals <larger than 7>, <between 5 and 7> and <lesser than 5> for high, medium and low voting respectively. However by looking at the data it is possible to note that the values are off by approximately a value of 1. This slight difference can be explained by the fact that even though subjects were made aware in the introduction to the questionnaire that the age damage of the drawings was not being modeled, this is not easily ignored by the subjects. For this reason the classification of high, medium and low is defined as <larger than 6>, <between 4 and 6> and <less than 4.5>.

Table 6 shows the average voting per question, the classification of the average voting, the expected classification and indicates whether the target was reached.

**Table 6.** Classification of the average per question (H: > 6, L: < 4.5, M: >= 4,5 and <= 6). Legend: x: reached, x-: reached below, xx: exceeded, n: not reached.

| | 1 | 2 | 3 | 4 | 5 | 6 | 7 | 8 | 9 | 10 | 11 | 12 | 13 | 14 | 15 | 16 | 17 | 18 | 19 | 20 | 21 | 22 | 23 | 24 | 25 | 26 | 27 | 28 |
|---|---|---|---|---|---|---|---|---|---|---|---|---|---|---|---|---|---|---|---|---|---|---|---|---|---|---|---|---|
| Average | 6,85 | 6,6 | 6,3 | 5,05 | 6,95 | 4,95 | 6,3 | 6,35 | 5,6 | 5,7 | 6,2 | 6,8 | 7,05 | 4,2 | 7 | 5,95 | 6,2 | 7,3 | 4,4 | 6,1 | 5,25 | 5,5 | 6 | 6,3 | 4,9 | 5,7 | 5,25 | 6,5 |
| Classification | H | H | H | M | H | M | H | H | M | M | H | H | H | L | H | M | H | H | L | H | M | M | M | H | M | M | M | H |
| Expected | H | H | H | M | MH | MH | MH | MH | H | H | M | ML | MH | MH | MH | M | ML | M | M | MH | MH | M | M | M | ML | M | M | M |
| Target reached | x | x | x | x | x- | x | x- | x- | x | x | xx | xx | x- | x | x- | x | x | xx | n | xx | x | x | x | x | x | x | x | xx |

By analyzing Table 6, we can see that 19 questions were classified as reached, 5 as reached below, 3 as exceeded and 1 not reached. These results confirm that the generated images can mimic the artistic style.

The results for the first sequence of questions (one to ten) support the expected result. The questions addressing the contour line characteristics (question one to five) are classified as high, with one of the best voting being obtained in question five, which indicates that the generated lines are considered to be quite similar to the original lines in the drawing. Unsurprisingly the worst result was obtained in question four as expected. Regarding the questions addressing lighting (question five to ten), some good results were achieved (questions seven and eight) with average results in the remaining questions. The inherent differences in the shapes being compared, and the degree of complexity of a shape influences the voting, for this reason, the simpler the shape being tested the higher voting is obtained. Question seven and eight represent simple shapes (leg and chest) whilst the remaining questions represent more complex shapes (several cloth folds), the variation in classification reflects closely the variation in shape complexity.

Questions eleven to sixteen represent the comparison between the the David statue rendering (Fig. 1-iv)) and the three drawings (Fig. 23, 24, 25). As expected this was the group of questions that obtained globally the highest classification of the three groups. It was also expected that the first two questions would have good results, and that the last two questions would have results above average, this was achieved. The classification obtained in question thirteen was a surprise since it was expected that it would receive a classification of Medium.

The difference between questions five and eleven was of 0,75 in their respective averages, with the lower classification occurring in question eleven. This corresponds to a difference of 7,5% of the maximum value, which can be considered as a significant variation. We note that there were also larger variations in voting between both questions with some of the subjects. This shows that it is difficult to separate the information of the contour lines from the information of the illumination and colour of the images. Both characteristics work together and one can affect the other.

From questions seventeen to twenty two the largest surprise was obtained in questions nineteen and twenty. Whilst question nineteen indicates that the values used when modeling the statue of the David are not adequate for the shape of the model used in this question, on the other hand question twenty shows that the simpler shape of the model creates little lighting variation, making it closer to the lighting used in the test drawing.

Questions twenty three to twenty eight received the expected classification.

Most of the comments made by subjects refer to problems encountered due to the age of the drawing and some about the quality of blow-up prints used in the questionnaire.

In the first five questions, addressing the characteristics of the line contours, despite of the overall high classification received, there where comments referring to differences in the colour or thickness of the lines in the different images. These comments show that, even when using similar shapes if one takes a closer look to smaller detail it is always possible to do better.

In the next five questions regarding illumination, the main criticism was the stronger illumination in some images. As expected, due to the differences in the scenes being represented, any region that is more lit is immediately noticed. However the simpler the image used in a question, less comments of this kind were received. This indicates that the illumination achieved the desired effect. In the questions regarding draping, one criticism was that the result resembled plasticine, this is somewhat similar to a reported result from an observational study on the use of NPR images in different contexts (e.g. comparing NPR with hand-drawn pen-and-ink scientific illustrations), which stated that the NPR result "looked like plastic", suggesting that the surface materials of the modeled object, not just the shading information should somehow be incorporated in the NPR system [Isenberg et al. 2006]. We believe that in our setting, further composition of the modeled style with a photo texture of a paper with similar grain to the drawing could mitigate this factor.

In the remaining questions, the produced comments referred to the same aspects as the previous ones. The comments received were also less, meaning that when comparing smaller parts of the images, there is always an aspect that can be reported, but when seeing the images in totality these small differences dilute and the global style is taken into account instead. This was hoped for when the questionnaire was designed.

## 5 Conclusion and future work

Nowadays the line between art and computer graphics becomes increasingly more blurred. For example, the computer can generate a preview image to help an artist to see how he can continue his work. The work presented here gives another digitally reproduced style to be used in that preview. The artistic style that António Campelo used in Fig. 23 was studied and its main characteristics were extracted and validated both from an image analysis viewpoint, and from a group of people. Our solution combines the strengths of existing image-space silhouette detection approaches with object-space solutions to fill in the gaps of contour line creation; depth information from Shadow Maps is weighted to change the tone of Phong shaded results, thus conveying the desired depth perception and highlights/shading contrast.

The results of the questionnaire showed that the desired effect was achieved, but it also showed the difficulties inherent in evaluating the work. Even though subjects were previously briefed on the specific characteristics that were being evaluated, abstracting from the evaluation the paper medium and damage is difficult and still affects the comparisons. However the obtained results were within the target values and allowed to complement the previous image-based evaluation and thus validate the obtained results.

The results also support that using parts of images for comparison rather than just the full images appears to be a promising way to overcome the problem that a direct comparison between images is not possible since the original models cannot be laser scanned, photographed or are fictional.

As a continuation of this work, we would like to model the bi-tonal shading present in Fig. 24 perhaps with a toon-shading approach, we would also like to apply/compose the modeled style to a model of paper or canvas where age/discoloration and damage of the works could also be simulated. Another possible direction of research would be to reduce or control the quantity of lines in a region. For example, the lines in the David model (Fig. 11-e)) hair region could be reduced; however this operation should be done carefully in order to avoid the removal of visually important lines, such as the exterior contour lines. The drawings main characteristics were considered global, however it could be interesting to study changes in these characteristics in different sub regions, which could produce interesting images. Finally the imperfections that may exist in a work could be studied. A flaw in a line or wrong colour are characteristics that could exist in a given work or be part of the actual style.

Finally our NPR tool (Fig. 17) can render/apply our António Campelo Alegoria da Prudência modeled style on a reduced model of the David statue (of 1 million triangles) at around 5 frames per second with a Core 2 Duo T9400 2.53GHz laptop with 3GB RAM and equipped with a GeForce 9700 M GT graphics card.

## 6 Acknowledgements

This work was supported by national funds through FCT - Fundação para a Ciência e Tecnologia - with reference UID/CEC/50021/2013.

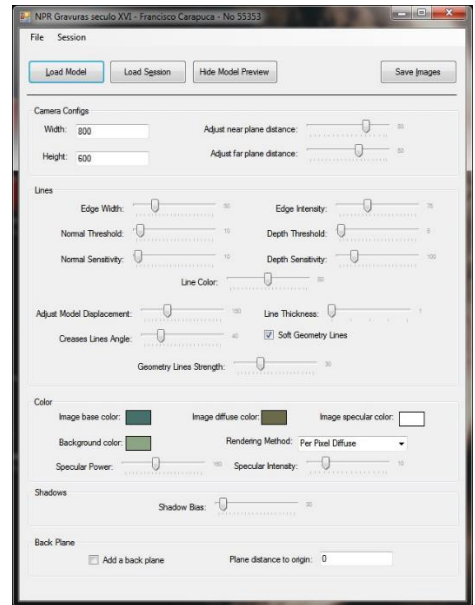

**Figure 17.** NPR parameter tuning.

Figures 23/1-i), 24, 25 were made available by Matrizpix (www.matrizpix.imc-ip.pt) and has the copyright of Instituto dos Museus e da Conservação IMC/MC. We would like to thank the reviewers and Bruno Araújo for their comments, the Museu Nacional de Arte Antiga (MNAA) museum, the MNAA curator Alexandra Markl, Chuva Vasco and Margarida Sardinha for invaluable art input on the work, and finally Stanford University for providing 3D models.

**Appendix A – Results**

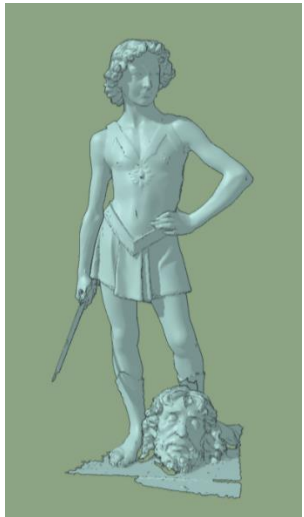
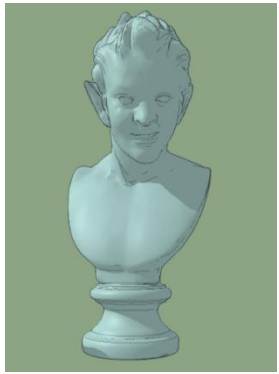
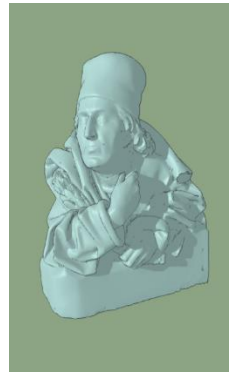
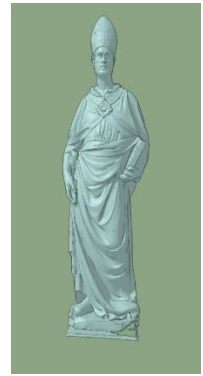
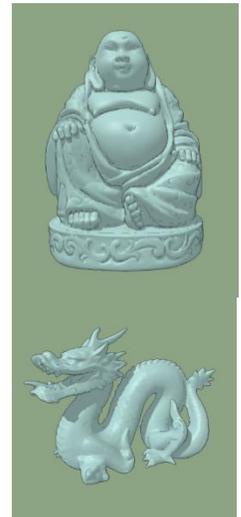

**Figure 19.** Statue 2    **Figure 18.** Statue 1    **Figure 20.** Statue 3    **Figure 21.** Statue 4    **Figure 22.** Statue 5&6

## Appendix B - Original Drawings

Images distributed by the Matrix Pix (htttp://www.matrizpix.imc-ip.pt/, division of the Portuguese Institute of Museums and Conservation (IMC). We note that the provided images are of low resolution, and that the IMC is not responsible for the quality of printed images in these conditions.

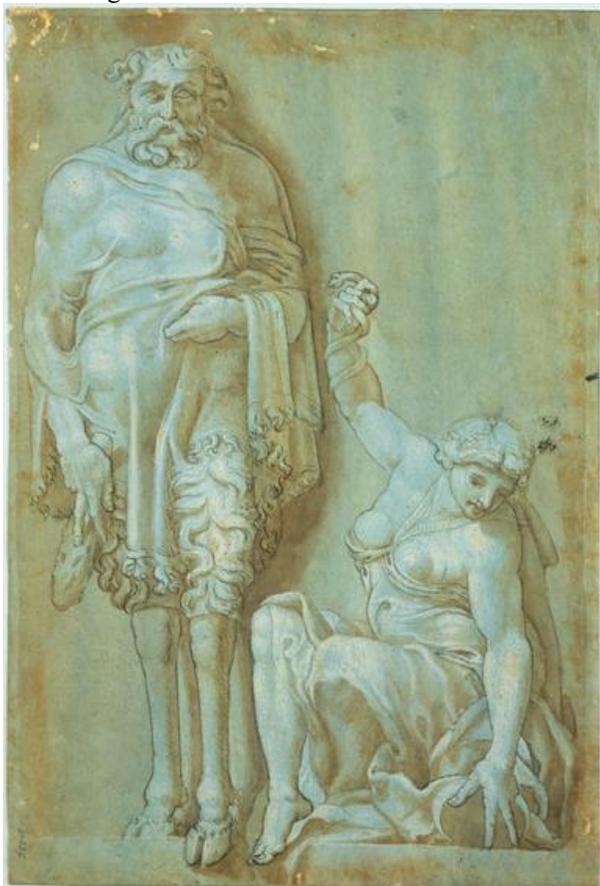

**Figure 23.** "Alegoria à Prudência" from António Campelo in the sixteenth century. Property of the Portuguese National Museum of Ancient Art. Photographed by Carlos Monteiro, 1994. Copyright © IMC / MC. Dimentions: H.32,7 x W.22,9 cm. Technical Information: Quill pen and Sepia wash highlighted in white gouache on blue paper preparation.

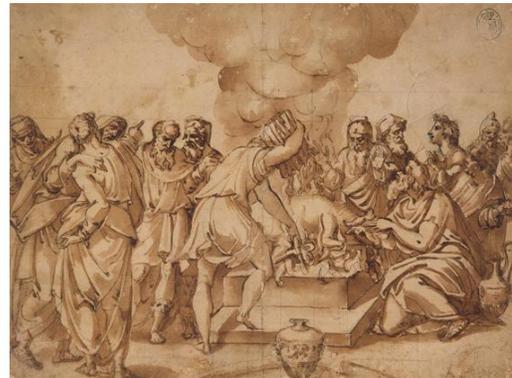

**Figure 24.** "Sacrifício Pagão" from António Campelo in the sixteenth century. Property of the Portuguese National Museum of Ancient Art. Photographed by José Pessoa, 1999. Copyright © IMC / MC. Dimensions: H.36,5 x W.48,5 cm. Technical Information: Quill pen and Sepia wash with a pencil grid.

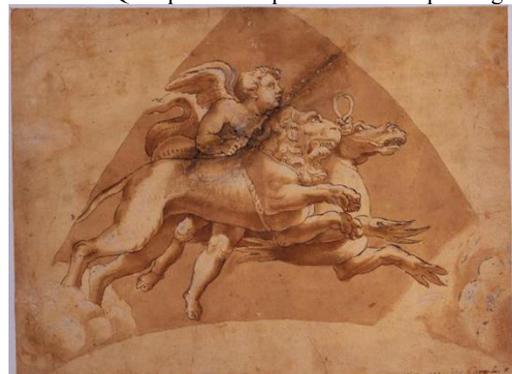

**Figure 25.** "Alegoria à Força" from António Campelo in the sixteenth century. Property of the Portuguese National Museum of Ancient Art. Photographed by Arnaldo Soares, 1993. Copyright © IMC / MC. Dimensions: H.20,3 x W.27,2 cm. Technical Information: Strokes and Sepia wash, highlighted in white (oxidated), light touches of sanguínea.